\documentclass[12pt]{iopart}
\usepackage{graphicx}
\usepackage{amssymb}
\usepackage{cite}

\def\abs#1{\vert #1 \vert}

\begin{document}
\vspace*{-3.2pc}
\title[Finite-temperature ordering in a two-dimensional
highly frustrated spin model]
{Finite-temperature ordering in a two-dimensional
highly frustrated spin model}
\author{A Honecker$^1$,
D C Cabra$^2$, H-U Everts$^3$, P Pujol$^4$ and F Stauffer$^{2,5}$}
\address{$^1$ Institut f\"ur Theoretische Physik,
 Georg-August-Universit\"at G\"ottingen, Friedrich-Hund-Platz 1,
 37077 G\"ottingen, Germany}
\address{$^2$ Universit\'{e} Louis Pasteur,
  Laboratoire de Physique Th\'{e}orique,
  67084 Strasbourg, C\'edex, France}
\address{$^3$ Institut f\"ur Theoretische Physik,
  Leibniz Universit\"at Hannover, Appelstra{\ss}e 2,
  30167 Hannover, Germany}
\address{$^4$ Laboratoire de Physique,
      ENS Lyon, 46 All\'ee d'Italie, 69364 Lyon C\'edex 07, France}
\address{$^5$ Institut f\"ur Theoretische Physik, Universit\"at zu K\"oln,
	Z\"ulpicher Stra{\ss}e 77, 50937 K\"oln, Germany}

\ead{ahoneck@uni-goettingen.de}

\begin{abstract}
We investigate the classical counterpart of an effective Hamiltonian for a
strongly trimerized kagom\'e lattice. Although the Hamiltonian only has a
discrete symmetry, the classical groundstate manifold has a continuous
global rotational symmetry. Two cases should be distinguished for the sign
of the exchange constant. In one case, the groundstate has a 120$^\circ$
spin structure. To determine the transition temperature, we perform
Monte-Carlo simulations and measure specific heat, the order parameter as
well as the associated Binder cumulant. In the other case, the classical
groundstates are macroscopically degenerate. A thermal order-by-disorder
mechanism is predicted to select another 120$^\circ$ spin-structure. A
finite but very small transition temperature is detected by Monte-Carlo
simulations using the exchange method.
\end{abstract}

\pacs{05.10.Ln,	
      64.60.Cn, 
      75.40.Cx} 
\vspace*{0.8pc}
{\small\rm
\begin{indented}
\item[] version of 8 September 2006
\vspace*{0.2pc}
\item[] to appear in: {\it \JPCM} (Proceedings of HFM2006, Osaka)
\end{indented}
}

\section{Introduction}

Highly frustrated magnets are a fascinating area of research with
many challenges and surprises \cite{SRFB04,Diep05}. One exotic
case is the spin-1/2 Heisenberg antiferromagnet on the kagom\'e
lattice, where in particular an unusually large number of low-lying
singlets was observed numerically \cite{lech97,waldt98}. An
effective Hamiltonian approach to the strongly trimerized kagom\'e lattice
\cite{sub95} was then successful in explaining
the unusual properties of the homogeneous lattice \cite{mila98}.

Interest in this situation has been renewed recently due to the
suggestion that strongly trimerized kagom\'e lattices can be
realized by fermionic quantum gases in optical lattices
\cite{SBCEFL04,FehrmannThesis}. For two fermions per triangle,
one obtains the aforementioned effective Hamiltonian on a
triangular lattice \cite{sub95}, but without the original
magnetic degrees of freedom. This effective Hamiltonian
also describes the spin-1/2 Heisenberg antiferromagnet on the
trimerized kagom\'e lattice at one third of the saturation magnetization
\cite{CGHHPRSS}. Furthermore, this model shares some features
with pure orbital models on the square lattice \cite{NuFr05,DBM05},
but it is substantially more frustrated than these models.

Numerical studies of the effective quantum Hamiltonian
\cite{DEHFSL05,DFEBSL05,CEHPSprep}
provide evidence for an ordered groundstate and a possible
finite-temperature ordering transition. Since the Hamiltonian only has
discrete symmetries, one expects indeed a finite-temperature
phase transition if the groundstate is ordered.
Furthermore, quantum fluctuations should be unimportant for the
generic properties of such a phase transition, which motivates
us to study the classical counterpart of the model at finite temperatures.

\section{Model and symmetries}

\begin{figure}[tb]
\begin{center}
\includegraphics[width=0.35\columnwidth]{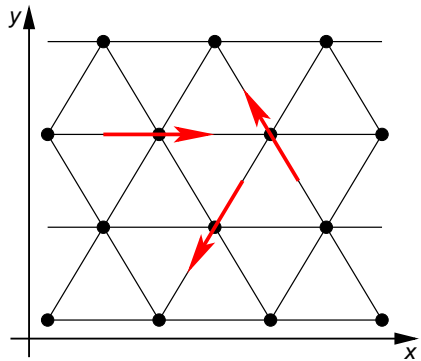}
\end{center}
\caption{Part of the triangular lattice. Arrows indicate the direction
of the unit vectors $\vec{e}_{i;\langle i,j\rangle}$ entering the
Hamiltonian (\ref{eqH}).
\label{fig:LatHam}
}
\end{figure}

In this paper we study a Hamiltonian which is given in terms
of spins $\vec{S}_i$ by
\begin{equation}
H = J \, \sum_{\langle i,j\rangle}
 \left(2 \, \vec{e}_{i;\langle i,j\rangle} \cdot \vec{S}_i\right) \,
 \left(2 \, \vec{e}_{j;\langle i,j\rangle} \cdot \vec{S}_j\right) \, .
\label{eqH}
\end{equation}
The sum runs over the bonds of nearest neighbours $\langle i,j\rangle$
of a triangular lattice with $N$ sites.
For each bond, only certain projections of the spins $\vec{S}_i$
on unit vectors $\vec{e}_{i;\langle i,j\rangle}$ enter the interaction.
These directions are
sketched in Fig.\ \ref{fig:LatHam}. Note that these directions depend
both on the bond $\langle i,j\rangle$ and the corresponding end $i$ or $j$
such that three different projections of
each spin $\vec{S}_i$ enter the interaction with its six nearest neighbours.

In the derivation from the trimerized kagom\'e lattice \cite{sub95,SBCEFL04,Z05},
the $\vec{S}_i$ are pseudo-spin operators acting on the two
chiralities on each triangle and should therefore be considered as quantum
spin-1/2 operators. Here we will treat the $\vec{S}_i$ as classical
unit vectors. Since only the $x$- and $y$-components enter the
Hamiltonian (\ref{eqH}), one may take the $\vec{S}_i$ as `planar' two-component
vectors. On the other hand, in the quantum case commutation relations
dictate the presence of the $z$-component as well, such that taking the
$\vec{S}_i$ as `spherical' three-component vectors is another natural choice
\cite{CEHPSprep}. Here we will compare both choices and thus assess qualitatively
the effect of omitting the $z$-components.

The internal symmetries of the Hamiltonian (\ref{eqH}) constitute the
dihedral group $D_6$, {\it i.e.}, the symmetry group of a regular hexagon.
Some of its elements consist of a simultaneous transformation of the spins
$\vec{S}_i$ and the lattice \cite{CEHPSprep}. Since these symmetries are
only discrete, a finite-temperature phase transition is allowed above an
ordered groundstate.

The derivation from a spin model \cite{sub95,Z05} yields a positive $J>0$,
while it may also be possible to realize $J<0$ \cite{FehrmannThesis} in
a Fermi gas in an optical lattice \cite{SBCEFL04}. In the following we will
first discuss the case of a negative exchange constant $J<0$ and then
turn to the case of a positive exchange constant $J>0$. The second case
is more interesting, but also turns out to be more difficult to handle.

\section{Negative exchange constant}

For $J<0$ there is a one-parameter family of ordered groundstates,
$\vec{S}_i = (\cos(\phi_i),\,\sin(\phi_i),\,0)$ with $\phi_a = \theta$,
$\phi_b = \theta+2\,\pi/3$ and $\phi_c = \theta-2\,\pi/3$ on the three
sublattices $a$, $b$, and $c$, respectively ($120^{\circ}$ N\'eel order).
The energy of these states, $E^{J<0} = 6\,J\,N$, is independent of
$\theta$. Computing the free energy $\mathcal{F}^{J<0}(\theta)$ by
including the effect of Gaussian fluctuations we find that
$\mathcal{F}^{J<0}(\theta)$ has minima at $\theta = (2\,n+1)\,\pi/6$,
$n = 0,\,1,\, \dots,\,5$ \cite{CEHPSprep}. This implies that the above
$120^{\circ}$ N\'eel structures lock in at these angles.

\begin{figure}[tb]
\begin{center}
\includegraphics[width=\columnwidth]{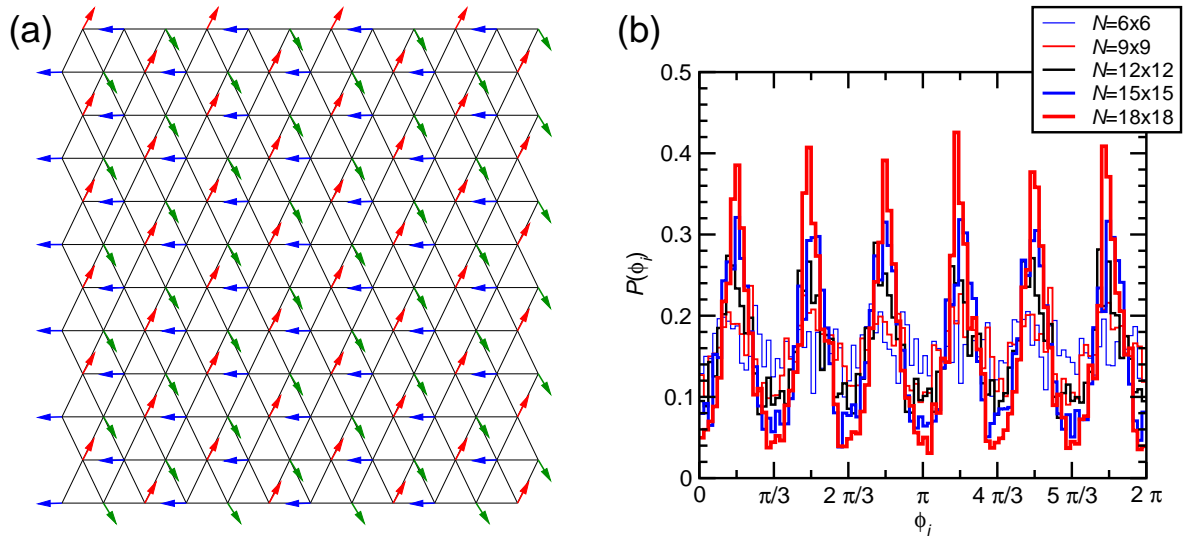}
\end{center}
\caption{MC results for $J<0$, $T = 10^{-3}\,\abs{J}$,
and planar spins. (a) Snapshot of a configuration
on a $12\times 12$ lattice.
Periodic boundary conditions are imposed at the edges.
(b) Histogram of angles $\phi_i$, averaged over 1000
configurations.
\label{fig:CFGneg}
}
\end{figure}

We have performed Monte-Carlo (MC) simulations for $J<0$ using a standard
single-spin flip Metropolis algorithm \cite{LB00}.
A snapshot of a low-temperature configuration on an
$N=12\times 12$ lattice is shown in Fig.\ \ref{fig:CFGneg}(a).
The $120^{\circ}$ ordering is clearly seen in such snapshots.
Fig.\ \ref{fig:CFGneg}(b) shows histograms of the angles $\phi_i$.
One observes that with increasing lattice size $N$, pronounced maxima
emerge in the probability $P(\phi_i)$ to observe an angle $\phi_i$
at the predicted lock-in values $\theta = (2\,n+1)\,\pi/6$.

\begin{figure}[tb]
\begin{center}
\includegraphics[width=0.65 \columnwidth]{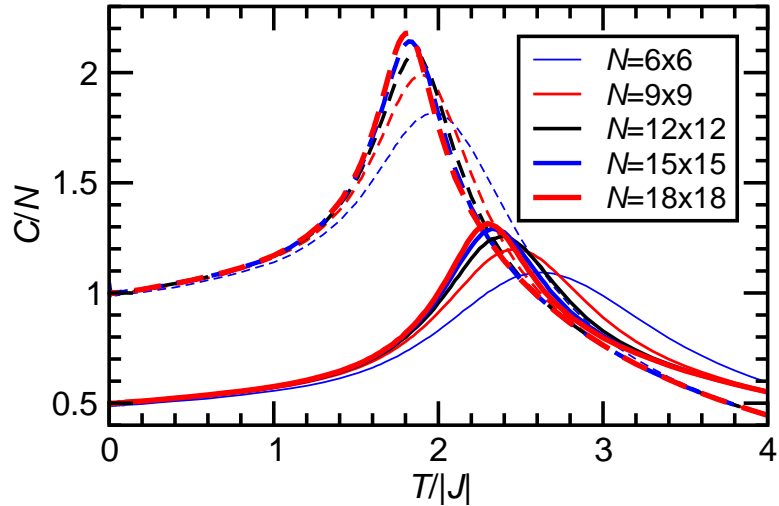}
\end{center}
\caption{MC results for the specific heat $C$ for $J<0$.
Full lines are for planar spins; dashed lines for spherical spins.
Increasing line widths denote increasing system sizes $N$.
Error bars are negligible on the scale of the figure.
\label{fig:Cneg}
}
\end{figure}

Thermodynamic quantities have been computed by averaging over at least 100
independent MC simulations. Each simulation was started at high
temperatures and slowly cooled to lower temperatures in order to minimize
equilibration times. A first quantity, namely the specific heat $C$,
is shown in Fig.\ \ref{fig:Cneg}. There is a maximum in $C$ at
$T \approx 1.8\,\abs{J}$ for spherical spins, and for planar spins at a
higher temperature $T\approx 2.2\,\abs{J}$. The fact that the value of
$C$ around the maximum increases with $N$ indicates a phase transition.
For $T \to 0$, the equipartition theorem predicts a contribution $1/2$
to the specific heat per transverse degree of freedom. Indeed, the
low-temperature results are very close to $C/N = 1/2$ and $1$ for planar
and spherical spins, respectively.

\begin{figure}[tb]
\begin{center}
\includegraphics[width=0.65 \columnwidth]{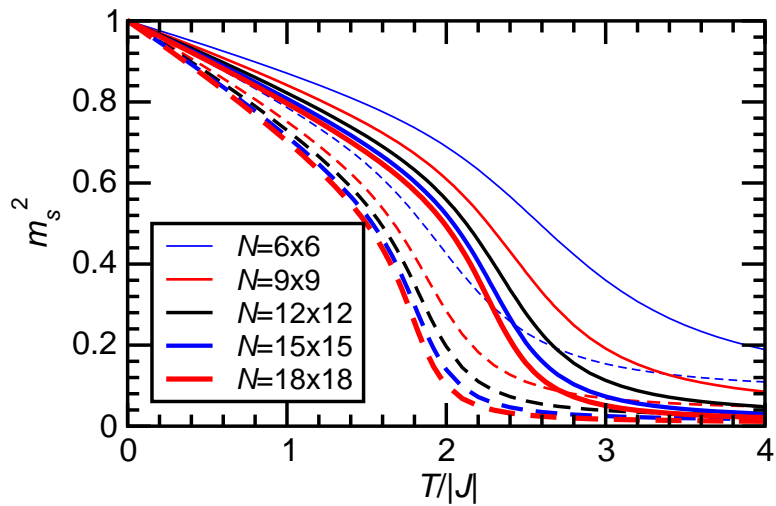}
\end{center}
\caption{MC results for the square of the sublattice
magnetization $m_s^2$ for $J<0$.
Full lines are for planar spins; dashed lines for spherical spins.
Increasing line widths denote increasing $N$.
Error bars are negligible on the scale of the figure.
\label{fig:Mneg}
}
\end{figure}

In order to quantify the expected
order, we introduce the sublattice order parameter
\begin{equation}
\vec{M}_s = {3 \over N} \sum_{i \in {\cal L}} \vec{S}_i \, ,
\label{defMsubl}
\end{equation}
where the sum runs over one of the three sublattices ${\cal L}$
of the triangular lattice. Fig.\ \ref{fig:Mneg} plots the square
of this sublattice order parameter
\begin{equation}
m_s^2 = \left\langle \vec{M}_s^2 \right\rangle \, ,
\label{defMs}
\end{equation}
which is a scalar quantity and expected to be non-zero in a
three-sublattice ordered state. Note, however, that the order
parameter (\ref{defMsubl}) is insensitive to the mutual orientation
of spins on the three sublattices.
In Fig.\ \ref{fig:Mneg} we observe that $m_s^2$ converges to
zero with $N \to \infty$ at high temperatures, while a non-zero
value persists at lower temperatures, consistent with a
phase transition around $T\approx 2\,\abs{J}$, as already
indicated by the specific heat. Furthermore, the sublattice order
again points to a higher transition temperature for planar spins than for
spherical spins. It should also be noted that there are noticeable
quantitative differences between the values of $m_s^2$ for
planar and spherical spins over the entire temperature range.
Just for $T \to 0$ both planar and spherical spins yield $m_s^2 \to 1$,
as expected for a perfectly ordered groundstate.

In order to determine the transition temperature $T_c$ more accurately,
we use the `Binder cumulant' \cite{Binder81,LB00} associated to
the order parameter (\ref{defMsubl}) via
\begin{equation}
U_4 = 1+ A - A \,
{\left\langle \vec{M}_s^4 \right\rangle \over
\left\langle \vec{M}_s^2 \right\rangle^2} \quad
{\rm with} \quad
A = \cases{1 & for planar spins, \cr
{3 \over 2} & for spherical spins. }
\label{defBinder}
\end{equation}
The constants in (\ref{defBinder}) are chosen such that $U_4=0$ for a
Gaussian distribution of the order parameter $P(\vec{M}_s) \propto
\exp\left(-c\,\vec{M}_s^2\right)$. Such a distribution is expected at high
temperatures, leading to $U_4 \to 0$ for $T \gg \abs{J}$. Conversely, a
perfectly ordered state yields $\left\langle \vec{M}_s^4 \right\rangle =
\left\langle \vec{M}_s^2 \right\rangle^2$ such that with the prefactors as
in (\ref{defBinder}) we find $U_4 = 1$. Hence, for an ordered state we
expect $U_4 \approx 1$ for $T < T_c$.

\begin{figure}[tb]
\begin{center}
\includegraphics[width=0.93 \columnwidth]{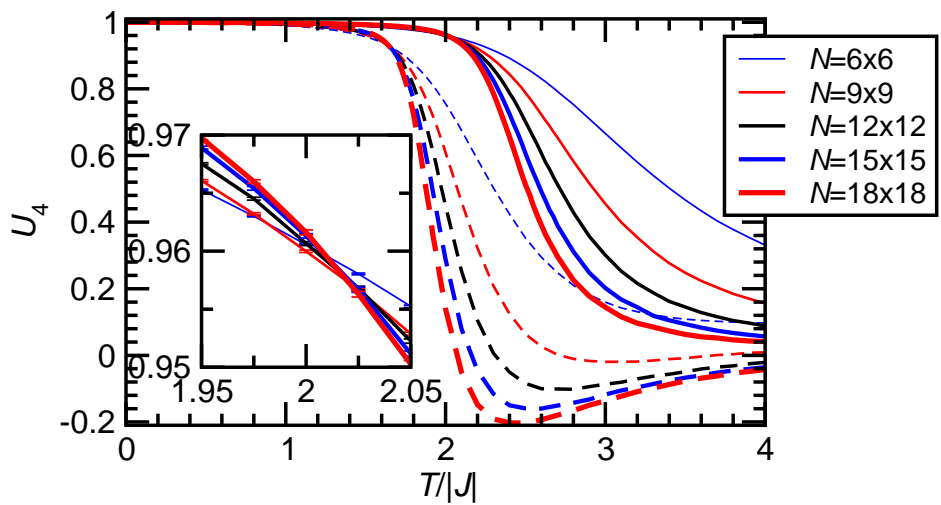}
\end{center}
\caption{Main panel:
MC results for the Binder cumulant for $J<0$.
Full lines are for planar spins; dashed lines for spherical spins.
Increasing line widths denote increasing $N$.
Error bars are negligible on the scale of the figure.
Inset: Binder cumulant for planar spins close to the
critical temperature.
\label{fig:Uneg}
}
\end{figure}

Fig.\ \ref{fig:Uneg} shows our results for the Binder cumulant, as defined
in (\ref{defBinder}). The transition temperature $T_c$ can be estimated
from the crossings of the Binder cumulants at different sizes $N$
\cite{Binder81,LB00}, which are shown for planar spins in the inset of
Fig.\ \ref{fig:Uneg}. Since we have not been aiming at high precision, we
cannot perform a finite-size extrapolation. Nevertheless, we obtain a
rough estimate
\begin{equation}
{T_c^{\rm planar}} \approx 2 \, {\abs{J}}\, ,
\label{TcNegPlanar}
\end{equation}
with an error on the order of a few percent. The corresponding
value for spherical spins is estimated as
${T_c^{\rm spherical}} \approx 1.57 \, {\abs{J}}$ \cite{CEHPSprep}.
We therefore conclude that out-of-plane fluctuations reduce $T_c$
by about $20\%$.

\section{Positive exchange constant}

As in the case $J<0$, there is a one-parameter family of $120^{\circ}$
N\'eel ordered groundstates with energy $-3\,J\,N$. They differ from the
states found for $J<0$ by an interchange of the spin directions on the
$b$- and $c$-sublattices. The contribution of Gaussian fluctuations around
these states yields a free energy $\mathcal{F}^{J>0}(\theta)$ which has
minima at $\theta = \pi \,n/3$, $n = 0,\,1,\, \dots,\,5$. Hence the N\'eel
structure locks in at these angles for $J>0$. Surprisingly, an inspection
of all states of finite cells of the lattice, in which the mutual angles
between pairs of spins are multiples of $2\,\pi/3$ \cite{CEHPSprep},
reveals that there is a macroscopic number of groundstates in addition to
the $120^{\circ}$ N\'eel states, {\it i.e.}, the number of ground states
grows exponentially with $N$. The $120^{\circ}$ N\'{e}el state has $N/3$
soft modes, all other groundstates have fewer soft modes \cite{CEHPSprep}.
At low but finite temperatures the $120^{\circ}$ N\'eel state will
therefore be selected by a thermal order-by-disorder mechanism \cite{OBD}.

\begin{figure}[tb]
\begin{center}
\includegraphics[width=0.5 \columnwidth]{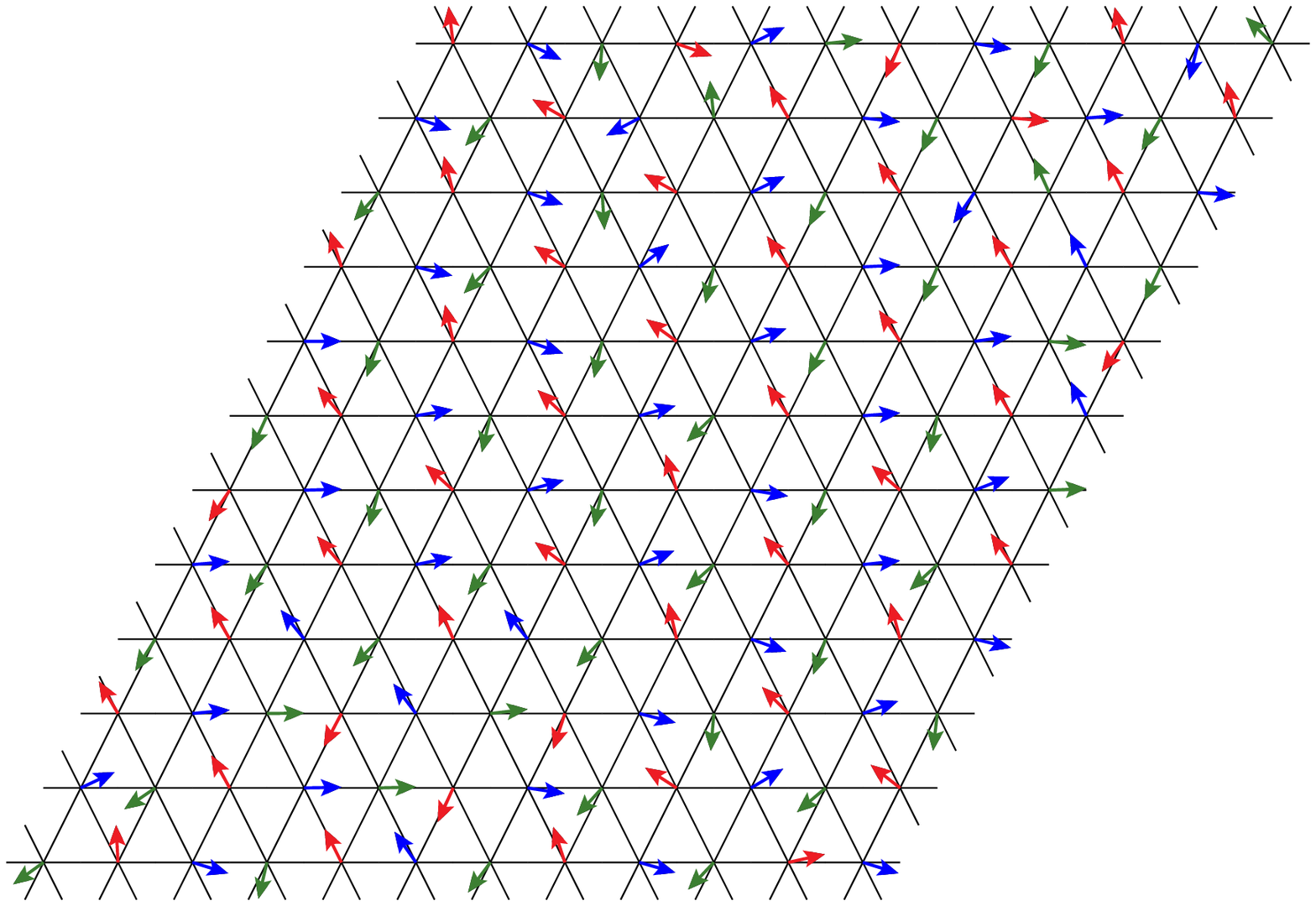}
\hspace*{-0.12 \columnwidth}
\includegraphics[width=0.5 \columnwidth]{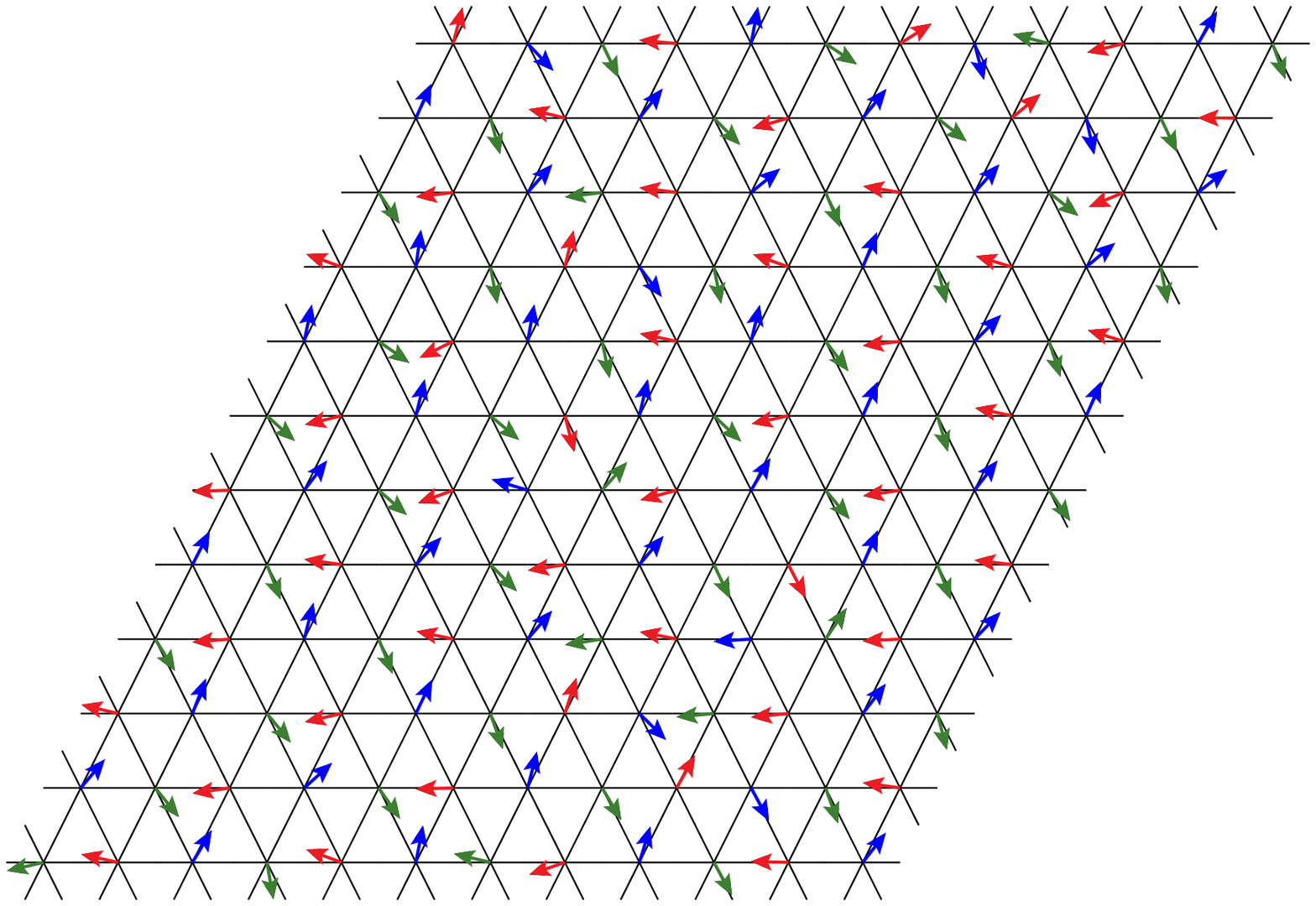}
\end{center}
\caption{Snapshots of configurations generated by the exchange
method at $T \approx 10^{-2} \, J$
on a $12\times 12$ lattice for planar spins and $J>0$.
Periodic boundary conditions are imposed at the edges.
\label{fig:CFGpos}
}
\end{figure}

The large number of groundstates which are separated by an energy barrier
renders it extremely difficult to thermalize a simple MC simulation at
sufficiently low temperatures. We have therefore performed exchange MC
simulations (also known as `parallel tempering') \cite{HN96,eM98} for
$J>0$, using a parallel implementation based on MPI. 96 replicas were
distributed over the temperature range $T/J = 0.01, \ldots, 0.7$ in a
manner to ensure an acceptance rate for exchange moves of at least 70\%.
Statistical analysis was performed by binning the time series at each
temperature. Fig.~\ref{fig:CFGpos} shows snapshots of two low-temperature
configurations with $N=12\times 12$ generated by the exchange method with
planar spins. In contrast to the case $J<0$, it is difficult to decide on
the basis of such snapshots if any order arises for $J>0$: there are
definitely large fluctuations including domain walls in the system. These
fluctuations are in fact a necessary ingredient of the order-by-disorder
mechanism. A careful quantitative analysis is therefore clearly needed.

\begin{figure}[tb]
\begin{center}
\includegraphics[width=0.65 \columnwidth]{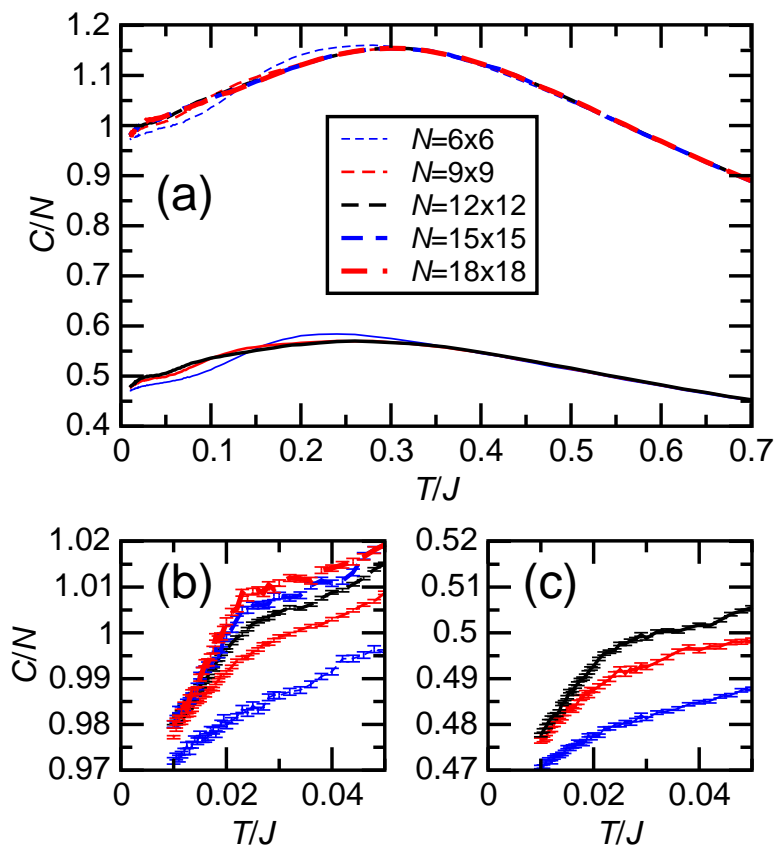}
\end{center}
\caption{Exchange MC results for the specific heat $C$ for $J>0$.
(a) Full lines are for planar spins; dashed lines for spherical spins.
Increasing line widths denote increasing system sizes $N$.
Error bars are on the order of the width of the lines in this panel.
The other panels show the
low-temperature parts of the results for spherical spins (b),
and planar spins (c).
The system sizes in panels (b) and (c) range from $N=6\times6$
(bottom) to $N=18\times18$ (panel (b), top),
and $N=12\times12$ (panel (c), top), respectively.
\label{fig:Cpos}
}
\end{figure}

We start with the specific heat $C$, shown in Fig.\ \ref{fig:Cpos}. There
is a broad maximum at $T \approx 0.25\,J$ for planar spins and $T \approx
0.3\,J$ for spherical spins (see Fig.\ \ref{fig:Cpos}(a)). However, this
maximum does not correspond to a phase transition, as one can infer from
the small finite-size effects. There is a second small peak in $C$ at a
lower temperature $T \approx 0.02 \, J$, see panels (b) and (c) of Fig.\
\ref{fig:Cpos}. Since this peak increases with $N$, it is consistent with
a phase transition. Note that at the lowest temperatures $C/N$ is clearly
smaller than $1/2$ and $1$ for planar and spherical spins, respectively.
Indeed, in-plane fluctuations around the $120^\circ$ state yield one
branch of soft modes, which we expect to contribute only $N/12$ to $C$
rather than $N/6$, as the other two branches. Thus, we expect $C/N = 5/12
= 0.41666\dots$ for planar spins and $C/N = 11/12 = 0.91666\dots$ for
spherical spins in the limit $T \to 0$. Our MC results tend in this
direction, but we have not reached sufficiently low temperatures to fully
verify this prediction. Lastly, we note that the difference between $C/N$
for planar and spherical spins is consistent with $1/2$ within error bars
only for $T \lesssim 0.02\, J$.

\begin{figure}[tb]
\begin{center}
\includegraphics[width=0.65 \columnwidth]{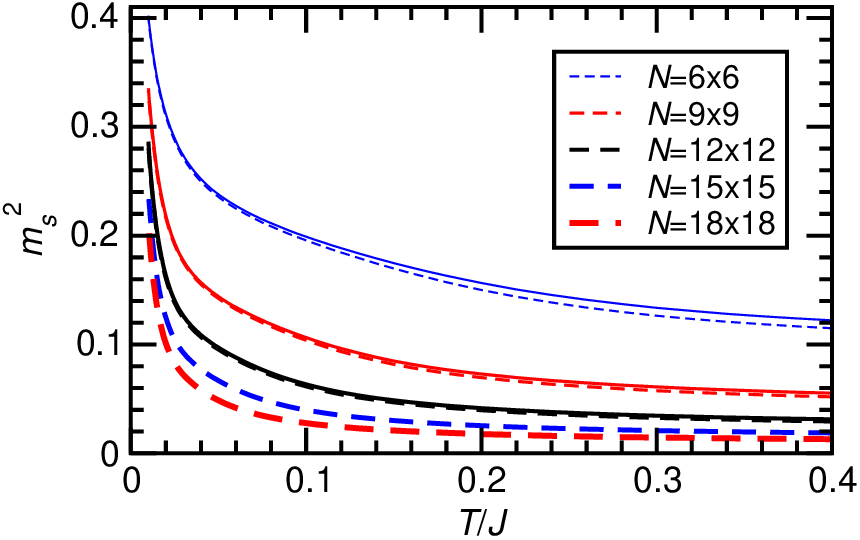}
\end{center}
\caption{Exchange MC results for the square of the sublattice
magnetization $m_s^2$ for $J>0$.
Full lines are for planar spins; dashed lines for spherical spins.
Increasing line widths denote increasing $N$.
Error bars are negligible on the scale of the figure.
\label{fig:Mpos}
}
\end{figure}

Next, in Fig.\ \ref{fig:Mpos} we show the square of the sublattice
magnetization (\ref{defMs}) for $J>0$. First, we observe that it remains
small for almost all temperatures and shoots up just at the left side of
Fig.\ \ref{fig:Mpos} where we measure a maximal value of $m_s^2 \approx
0.4$ for $T=J/100$ on the $N=6\times 6$ lattice. This is consistent with a
phase transition into a three-sublattice ordered state in the vicinity of
the low-temperature peak in the specific heat $C$. In marked difference
with the case $J<0$ (see Fig.\ \ref{fig:Mneg}), here we observe only small
differences between planar and spherical spins (at least for the sizes
where we have data for planar spins, {\it i.e.}, $N=6\times6$, $9\times9$
and $12\times12$).

\begin{figure}[tb]
\begin{center}
\includegraphics[width=0.98 \columnwidth]{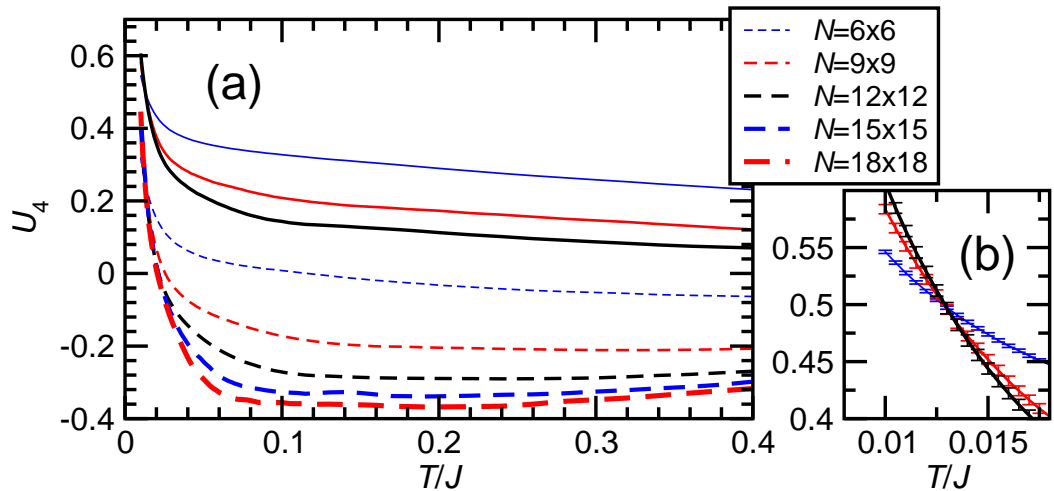}
\end{center}
\caption{Exchange MC results for the Binder cumulant for $J>0$.
(a) Full lines are for planar spins; dashed lines for spherical spins.
Increasing line widths denote increasing $N$.
Error bars are at most on the order of the width of the lines.
(b) Binder cumulant for planar spins close to the
critical temperature (\ref{TcPosPlanar}). Lines of increasing slope
are for increasing system sizes $N=6\times6$, $9\times 9$,
and $12\times12$.
\label{fig:Upos}
}
\end{figure}

Finally, Fig.\ \ref{fig:Upos} shows our results for the Binder cumulant,
as defined in (\ref{defBinder}). In contrast to the sublattice
magnetization $m_s^2$, the Binder cumulants for planar and spherical spins
are close to each other only for $T \lesssim 0.1\,J$, provided we scale to
the same prefactor $A$ in the definition (\ref{defBinder}). Again, our
best estimate for the transition temperature $T_c$ is obtained from the
crossings of the Binder cumulants at different sizes $N$ shown for planar
spins in the inset of Fig.\ \ref{fig:Upos}. The only meaningful crossings
are those of the $N=6\times6$ with the $N=9\times 9$ and $12\times12$
curves, respectively. This leads to a rough estimate
\begin{equation}
{T_c^{\rm planar}} \approx 0.0125 \, {J}\, .
\label{TcPosPlanar}
\end{equation}
This estimate is indistinguishable from the corresponding value for
spherical spins \cite{CEHPSprep}, as is expected since the spins lie
essentially all in the $x$-$y$-plane for such low temperatures.

\section{Conclusions and outlook}

In this paper,
we have investigated finite-temperature properties of the classical
version of the Hamiltonian (\ref{eqH}) on a triangular lattice.

For $J<0$, we have clear evidence for a low-temperature phase
with $120^\circ$ N\'eel order. The transition temperature is of
order $\abs{J}$ and can be determined with reasonable accuracy.
Planar and spherical spins yield qualitatively similar results,
but there are quantitative differences, in particular the
transition temperature (\ref{TcNegPlanar}) is higher for planar spins.
We have performed further simulations for spherical spins \cite{CEHPSprep}
in order to determine critical properties. On the one hand, so far
we have no evidence for a discontinuity at $T_c$, on the other hand the
assumption of a continuous phase transition yields very unusual
critical exponents \cite{CEHPSprep} which violate the hyperscaling
relation. Thus, the most plausible scenario may be a
weakly first-order transition.

For $J>0$, the groundstates are macroscopically degenerate \cite{CEHPSprep}.
A thermal order-by-disorder mechanism \cite{OBD} predicts the
selection of another $120^\circ$ ordered state. The corresponding
phase transition is just at the limits of detectability even with
the exchange MC method \cite{HN96,eM98}: we find an extremely
low transition temperature (\ref{TcPosPlanar})
which is two orders of magnitude smaller than the overall energy scale $J$.
In this case, the spins lie essentially in the $x$-$y$-plane in the
relevant temperature region such that we obtain quantitatively extremely
close results for planar and spherical spins in the vicinity of $T_c$,
apart from a constant offset in the specific heat.

The features observed e.g.\ in the specific heat for the classical variant
resemble those found in the original quantum model
\cite{DEHFSL05,DFEBSL05,CEHPSprep}. Since only much smaller systems are
numerically accessible for the quantum model, the results for the classical
variant are an important tool for understanding the quantum case.
In particular, the finite-temperature phase transitions should be universal
and may therefore be characterized in the classical model. However,
highly accurate data is needed for that purpose. Our result that spherical
and planar spins exhibit the same qualitative features for $J<0$, and that
for $J>0$ the $z$-component is even quantitatively very small in the
relevant temperature range, may be useful in this context. Namely,
one may choose planar spins for further simulations, thus
reducing the number of degrees of freedom to be updated.

\ack

Useful discussions with F.\ Mila are gratefully acknowledged. We are
indebted to the CECPV, ULP Strasbourg for allocation of CPU time on an
Itanium 2 cluster. This work has been supported in part by the European
Science Foundation through the Highly Frustrated Magnetism network.
Presentation at HFM2006 is supported by the Deutsche
Forschungsgemeinschaft through SFB 602.

\end{document}